\documentclass[conference]{IEEEtran}
\IEEEoverridecommandlockouts
\usepackage{cite}
\usepackage{amsmath,amssymb,amsfonts}
\usepackage{graphicx}
\usepackage{textcomp}
\usepackage{xcolor}
\usepackage{amsmath, amssymb}
\usepackage{algorithm}
\usepackage{algpseudocode}
\usepackage{physics}
\usepackage{float} 
\usepackage{subcaption}
\usepackage{booktabs}
\usepackage{multirow}
\usepackage{wrapfig} 
\usepackage{array}    
\usepackage{booktabs} 
\usepackage{caption}  
\usepackage{adjustbox}
\usepackage{fancyhdr}
\usepackage{bm}
\begin{document}

\fancypagestyle{firstpage}{
    \fancyhf{}
    \fancyfoot[C]{\footnotesize\itshape
    Accepted author manuscript at the IEEE International Conference on Communications, Control, and Computing Technologies for Smart Grids (IEEE SmartGridComm 2026). This version has been accepted for publication and may differ slightly from the final published version.}
    \renewcommand{\headrulewidth}{0pt}
    \renewcommand{\footrulewidth}{0pt}
}

\title{Inertia-Informed Federated Learning Control Framework for Distributed Smart Grid Resilience}

\author{
\IEEEauthorblockN{
Ibrahim Shahbaz,
Omar Al-Refai,
Eman Hammad}
\IEEEauthorblockA{\textit{iSTAR Lab, Texas A\&M University}\\
College Station, TX, USA\\
\{i.shahbaz, omaralrefai, eman.hammad\}@tamu.edu}
}

\maketitle
\thispagestyle{firstpage}

\begin{abstract}
Resilient-by-design smart grid control demands frameworks capable of maintaining stability under physical disturbances and communication failures, without reliance on centralized coordination. While Centralized Training Decentralized Execution (CTDE) enables a learning-based control paradigm at the grid edge, individually trained models fail to generalize across unseen fault contingencies and fall short of fully decentralized deployment. Federated learning (FL) restores generalization through collaborative training; however, standard aggregation strategies remain agnostic to the physical heterogeneity of synchronous generators. This work proposes Inertia-Informed Weighted FedAvg (IIWFedAvg), a physics-informed aggregation strategy that embeds generator inertia directly into global model fusion for transient stability control in transmission networks. The proposed framework further integrates interpretable Chebyshev Kolmogorov-Arnold Network (ChebyKAN)-based controllers, augmented with Rate-of-Change-of-Frequency (RoCoF) features to enhance dynamic response awareness. Evaluated on the IEEE 39-bus benchmark under full decentralized deployment, IIWFedAvg achieves a 75\% generalization success rate across unseen fault contingencies. It also surpasses the centralized baseline in two out of three stabilized faults, while delivering a 3× improvement in stabilization speed at zero centralized coordination overhead.
\end{abstract}

\begin{IEEEkeywords}
Federated learning, physics-informed aggregation, Kolmogorov-Arnold networks, transient stability, smart grid resilience, distributed control.
\end{IEEEkeywords}

\section{Introduction}
\label{sec:intro}
 
Modern power systems are evolving into cognitive cyber-physical infrastructures with pervasive sensing, computation, communication, and control to support reliable operation and high penetration of distributed energy resources. In this context, resilience has emerged as a core design objective for smart grid (SG) control, requiring the system to maintain stable, reliable, and secure operation under a wide range of disturbances, uncertainties, and component failures~\cite{PG_reselience_adverseries,impact_cyber_attacks}. A key challenge is the trade-off between centralized and decentralized control schemes: while centralized approaches exploit global information for near-optimal performance, they are vulnerable to communication failures and single points of failure; conversely, decentralized approaches improve robustness and scalability but often sacrifice performance due to limited system-wide visibility~\cite{farraj2016cyber}. This work addresses transient stability by proposing an inertia-informed centralized training and decentralized execution (CTDE) control framework that leverages federated learning (FL) to approximate centralized closed-loop actions under unseen fault contingencies in fully decentralized operation.

Recent advances in learning-based control, particularly reinforcement learning (RL) and FL, offer a promising pathway beyond the limitations of traditional control schemes~\cite{ahmadi2026comprehensive}. RL synthesizes non-linear control policies directly from simulation interaction data, making it well-suited to handle complex grid dynamics~\cite{RL_MG_Control}, yet its online closed-loop deployment remains constrained by the need for extensive offline episodic training, while the weak interpretability of learned policies further hinders trust and adoption in safety-critical settings~\cite{XAI_PS}. Complementarily, FL enables multiple agents to collaboratively train shared models without exchanging raw data, preserving privacy and reducing communication overhead~\cite{FedAvg}. The performance of FL-based control, however, is ultimately governed by the aggregation strategy that fuses locally trained models into a global model~\cite{weber2024combining}.
 
Despite extensive progress in FL, widely used aggregation strategies have largely been developed for generic machine-learning settings, where the primary focus is on mitigating statistical heterogeneity. Representative methods such as FedAvg~\cite{FedAvg}, and FedProx~\cite{fedprox} are designed to handle non-independent and identically distributed (non-IID) data distributions and class imbalance across clients. More recently, personalized and physics-informed FL variants have begun to incorporate domain knowledge either through modified local objectives or tailored model architectures~\cite{tan2022personalized}. Still, embedding physical properties directly into the aggregation strategy itself remains largely unaddressed in power system applications.
 
Beyond raw performance, the deployment of learning-based controllers in safety-critical infrastructure requires model interpretability~\cite{Henao2025AIi}. In prior work~\cite{GPIFNN}, a lightweight, inherently interpretable neural architecture was proposed that provides a foundational basis for physics-informed modeling of electrical systems. Building on this direction, Kolmogorov--Arnold Networks (KANs)~\cite{KANs} 
and their Chebyshev variants (ChebyKANs)~\cite{chebyKAN} approximate 
functions through learnable univariate basis functions, yielding inherently interpretable representations absent in conventional 
neural network architectures. This combination of expressiveness and transparency has positioned KANs as an emerging tool for power-system applications where interpretability is critical, ranging from physics-informed modeling of grid dynamics~\cite{PINNS_KANs} to the detection of cyber-attacks in automatic generation control loops~\cite{KANs_for_Cyber_detection_AGC} and electric vehicle charging infrastructure~\cite{saber2025kolmogorov}.
 
This work is motivated by two critical observations in prior studies regarding the gap between offline training performance and online closed-loop reliability. In \cite{evaluate_KANS_paper}, individual KAN-based controllers were shown to achieve high accuracy during offline training; however, they often fail in online closed-loop stabilization under unseen contingencies. Additionally, empirical results indicated that controllers trained on actions of high-inertia generators exhibit improved transferability, providing a stabilizing effect even when deployed on lower-inertia generator buses. Furthermore, although standard FL aggregation (FedAvg) has demonstrated improved generalization across unseen disturbances \cite{FLC_Paper}, its effectiveness remains limited to moderate levels of decentralized deployment and does not achieve fully stable distributed control. These findings suggest that generic aggregation strategies fail to exploit fundamental physical properties, particularly inertia heterogeneity, that govern power system stability.

To address this gap, an \emph{Inertia-Informed Weighted Federated Averaging} (IIWFedAvg) strategy is proposed, which explicitly incorporates generator physical properties into the global model fusion process. By weighting global model updates according to each generator's inertia constant, the approach emphasizes the stabilizing contribution of high-inertia units while preserving the diverse operational knowledge of the overall system. This physics-informed aggregation, implemented through interpretable ChebyKAN-based controllers under a CTDE framework, enables stable closed-loop operation and improves generalization to out-of-distribution disturbances in fully decentralized controller deployment.

\section{Background}
\label{sec:back}
 KANs constitute a recent family of neural function approximators whose mathematical foundation is the Kolmogorov--Arnold superposition principle. The principle establishes that any continuous mapping $f:[0,1]^n \to \mathbb{R}$ admits an exact decomposition into a finite composition of uni-variate functions~\cite{KANs}:
\begin{equation*}
f(x_1,\ldots,x_n)
= \sum_{q=1}^{2n+1} \Phi_q\!\left( \sum_{p=1}^{n} \phi_{q,p}(x_p) \right),
\end{equation*}
where $\phi_{q,p}:[0,1]\to\mathbb{R}$ and $\Phi_q:\mathbb{R}\to\mathbb{R}$ are continuous scalar functions. The per-edge transformation is implemented as a finite linear combination of basis functions, $\phi(x)=\sum_{i=0}^{n} c_i B_i(x)$, with the coefficients $\{c_i\}$ optimized during training. Since each $\phi$ is a univariate map over a bounded interval, it can be visualized and approximated by a closed-form symbolic expression, giving the trained network a layer of interpretability that conventional architectures do not provide.

The ChebyKAN variant retains the edge-wise functional structure of a KAN but substitutes the spline basis with a Chebyshev polynomial expansion~\cite{chebyKAN}, casting each edge as: $\phi(x)=\sum_{n=0}^{d} c_n T_n(x)$, where $d$ denotes the truncation order, $\{c_n\}$ are trainable coefficients, and $T_n$ refers to the Chebyshev polynomials of the first kind, defined on $[-1,1]$ by $T_n(x)=\cos\!\left(n\arccos(x)\right)$. Equivalently, $T_n(x)$ can be generated through the recurrence
\begin{equation*}
T_0(x)=1, \quad T_1(x)=x, \quad T_n(x)=2xT_{n-1}(x)-T_{n-2}(x),
\end{equation*}
for $n \ge 2$, which provides a numerically efficient method for evaluating the polynomials at arbitrary order. 

ChebyKANs exploit the orthogonality and global approximation properties 
of Chebyshev polynomials, yielding greater numerical stability and 
expressiveness per parameter than spline-based KANs, while preserving 
their inherent interpretability. ChebyKANs serve as the local function 
approximators within the FL control framework introduced in 
Section~\ref{sec:FLC}.

\section{System Model}
\label{sec:setup}

\begin{figure*}[t]
\centering
\includegraphics[width=0.95\linewidth]{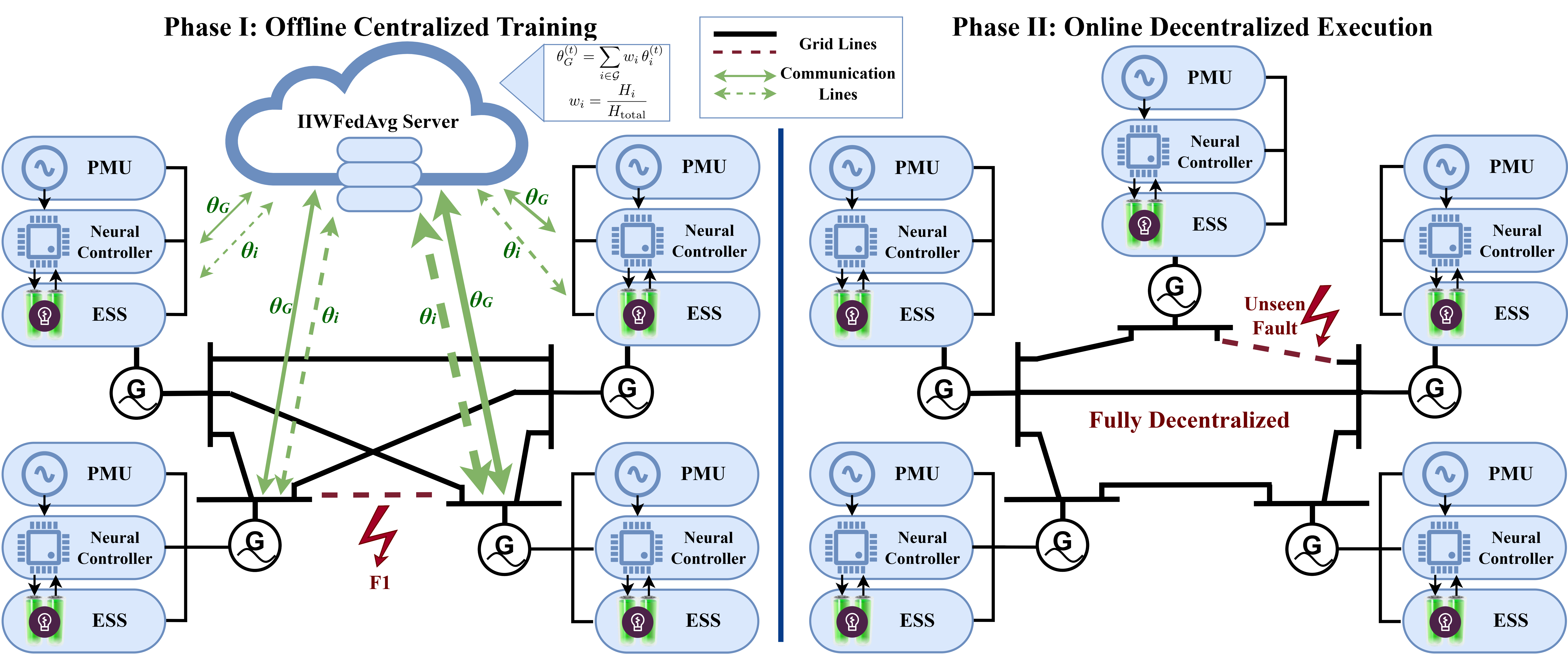}
\caption{Proposed inertia-informed federated learning control framework.}
\label{fig:FL_SG}
\end{figure*}

The transient stability control of SG transmission networks is formulated as a CTDE multi-agent framework, as illustrated in Fig.~\ref{fig:FL_SG}. Let the system be composed of $N$ synchronous generators, each associated with a local intelligent agent comprising: (i) a Phasor Measurement Unit (PMU) providing real-time local measurements, (ii) a fast-acting energy storage system (ESS) enabling bidirectional active power exchange, and (iii) a local neural controller.

\subsection{System Dynamics}
The electromechanical behavior of the interconnected generators is characterized by the classical swing equation~\cite{PS_Analysis}, which captures the evolution of each machine's rotor angle $\delta_i$ and angular frequency $\omega_i$. For generator~$i$, $\forall i \in \{1, \dots, N\}$, the dynamics are expressed as $\dot{\delta}_i = \omega_i$ and
\begin{equation}\label{eqn: swing 2}
H_i \dot{\omega}_i = -D_i \omega_i + (P_{m,i} - P_{e,i}),
\end{equation}
where~$H_i$ denotes the inertia constant (proportional to the stored kinetic energy of the rotating mass), $D_i$~is the damping coefficient, $P_{m,i}$~is the mechanical power input, and~$P_{e,i}$ is the electrical power output. The inertia constant~$H_i$ is of particular significance in this work, as it governs the sensitivity of generator~$i$ to torque imbalances: a larger~$H_i$ results in a slower, more stable frequency response that is inherently less susceptible to transient oscillations. The electrical power exchanged through the transmission network couples all generators and is given by:
\begin{equation*}\label{eqn: electrical power}
P_{e,i} = \sum_{k=1}^{N} |E_i|\,|E_k| \left[ G_{ik} \cos(\delta_i - \delta_k) + B_{ik} \sin(\delta_i - \delta_k) \right], 
\end{equation*}
where~$|E_i|$ and~$|E_k|$ are the internal voltage magnitudes of generators~$i$ and~$k$, respectively, and~$G_{ik} = G_{ki} \geq 0$, $B_{ik} = B_{ki} > 0$ are the conductance and susceptance elements obtained via Kron reduction~\cite{kron_reduction} of the full network admittance matrix to the generator-bus subspace.

\subsection{Baseline Control Architecture}
The baseline post-disturbance stabilization control mechanism considered in this work is the centralized parametric feedback linearization~(CPFL) controller~\cite{re_feedback_linearization}, which computes the auxiliary control action required at generator~$i$ bus as:
\begin{equation}\label{eqn: CPFL Pu}
P_{u,i} = -\left( P_{a,i} - P_{d,i}  \right),
\end{equation}
where~$P_{a,i} = P_{m,i} - P_{e,i}$ is the accelerating power. The decentralized PFL~(DPFL) component relies exclusively on local PMU observations and is defined as
\begin{equation}\label{eqn: DPFL Pu}
P_{d,i} = -\left( \alpha_i \omega_i + \beta_i (\delta_i - \delta_i^*) \right),
\end{equation}
where~$\alpha_i, \beta_i \geq 0$ are the frequency and phase stabilization gains, and~$\delta_i^*$ is the target rotor angle setpoint. The control signal~$P_{u,i}$ is designed to cancel the non-linear terms in~\eqref{eqn: swing 2}, thereby achieving feedback linearization of the local dynamics. Physically, $P_{u,i}$~commands the active power exchange of the co-located ESS: a negative value corresponds to power absorption (ESS charging), while a positive value corresponds to power injection (ESS discharging).

\section{Inertia-Informed Federated Learning Control}\label{FLC_framework}
\label{sec:FLC}

The proposed inertia-informed federated learning control framework, illustrated in Fig.~\ref{fig:FL_SG}, consists of~$N$ intelligent agents coordinated by a central FL server. Each agent hosts a local ChebyKAN-based neural controller deployed as an edge device at a generator bus alongside a fast-acting ESS. During training, each local controller transmits its learned parameter vector~$\theta_i$ to the FL server, which aggregates the individual policies into a global control model that encodes the collective generator dynamics. The updated global model is then broadcast back to all agents for the 
next learning round. Two key modifications over the FL control framework proposed in~\cite{FLC_Paper} are introduced in this section: (i)~a rate-of-change-of-frequency~(RoCoF) input augmentation that enriches the local neural controller's observability of remote system dynamics, and (ii)~an inertia-informed federated aggregation strategy that embeds swing-equation physics into the global model fusion.

\subsection{RoCoF-Augmented Local Neural Controller}

In prior work~\cite{FLC_Paper}, each local neural controller receives a three-dimensional input comprising the angular frequency deviation~$\omega_i$, the rotor angle deviation~$\Delta\delta_i = \delta_i - \delta_i^*$, and the simulation time index~$t$. While sufficient for single-fault training, this input space introduces a fundamental limitation: two physically distinct fault scenarios can produce identical local observations~$(\omega_i, \Delta\delta_i, t)$ while requiring different control actions, since the accelerating power~$P_{a,i}$ depends on the global configuration of remote rotor angles~$\{\delta_k\}_{k \neq i}$ through the electrical coupling in~$P_{e,i}$.

To address this, we augment the local input space with a new feature, the rate of change of frequency~(RoCoF), defined as:
\begin{equation}\label{eqn:rocof}
\dot{\omega}_i \approx \frac{\omega_i(t) - \omega_i(t - \Delta t)}{\Delta t},
\end{equation}
which is entirely computable from consecutive local PMU samples, without any inter-agent communication. From the swing equation~\eqref{eqn: swing 2}, the RoCoF satisfies
\begin{equation}\label{eqn:pa_rocof}
\dot{\omega}_i = \frac{1}{H_i}\left[-D_i \omega_i + P_{a,i}\right],
\end{equation}
revealing that~$\dot{\omega}_i$ encodes aggregated information of the influence of all remote generators on the local dynamics through the power flow coupling embedded in~$P_{a,i}$. This provides any neural controller with an implicit proxy for the unobserved global state at zero communication cost.

\subsection{Physics-Informed Inertia-Weighted Aggregation}

The standard Federated Averaging (FedAvg)~\cite{FedAvg} algorithm computes the FL global model as the unweighted mean~$\theta_G = \frac{1}{N}\sum_{i}\theta_i$, treating all generators identically regardless of their physical characteristics. This uniform treatment disregards the heterogeneous inertia structure of the power system, assigning equal influence to both low-inertia (which exhibit fast, oscillation-prone dynamics) and high-inertia generators.

The proposed \emph{Inertia-Informed Weighted FedAvg}~(IIWFedAvg) strategy addresses this limitation by deriving the aggregation weights directly from the swing equation~\eqref{eqn: swing 2}. Since inertia constants $H_i$, are known, time-invariant physical parameters of any commissioned synchronous generator, they provide a readily available and deterministic basis for aggregation weighting. A generator with a larger~$H_i$ possesses greater resistance to frequency deviations, producing a local control policy that is less sensitive to transient perturbations. We therefore weight each generator's contribution to the global model in proportion to its inertia, reflecting their stabilizing influence on system dynamics:
\begin{equation}\label{eqn: IIWFedAvg}
\theta_{G}^{(t)} = \sum_{i \in \mathcal{G}} w_i \, \theta_i^{(t)}, \qquad w_i = \frac{H_i}{H_{\mathrm{total}}}, \qquad \sum_{i \in \mathcal{G}} w_i = 1,
\end{equation}
where~$\mathcal{G} = \{G_1 ,G_i, \ldots, G_{N}\}$ denotes the set of~$N$ active generators, and~$H_{\mathrm{total}} = \sum_{j \in \mathcal{G}} H_j$. The weights~$\{w_i\}$ are fixed across all federated rounds, since the inertia constants are physical machine parameters that remain invariant during normal operation. IIWFedAvg ensures that generators with higher-inertia exert the greatest influence over the global control policy, while lower-inertia generators contribute proportionally less. This physics-informed weighting yields an aggregation strategy that is fully transparent, traceable to the swing equation, and deterministic.

\subsection{Approximated Control Action and Deployment}
Each local ChebyKAN controller $f_{\theta_i}$ is trained to approximate the nonlinear accelerating power $P_{a,i}$ from \eqref{eqn: CPFL Pu} using a four-dimensional local observation vector. Training is performed by minimizing the local loss function:
\begin{equation}\label{eqn:local_loss}
\mathcal{L}_{i}(\theta_{i}) = \left\| f_{\theta_i}(\omega_i, \Delta\delta_i, \dot{\omega}_i, t) - P_{a,i} \right\|^2.
\end{equation}
The approximated control action is:
\begin{equation}\label{eqn: FLC Pu}
\hat{P}_{u,i} = -\left( \hat{P}_{a,i} - P_{d,i} \right), \qquad 
\hat{P}_{a,i} = f_{\theta_i}(\omega_i, \Delta\delta_i, \dot{\omega}_i, t),
\end{equation}
constituting a decentralized learning-based feedback linearization that approximates the centralized control action  in~\eqref{eqn: CPFL Pu} using strictly local PMU measurements. The complete framework is summarized in \textbf{Algorithm~\ref{alg:FLC}}.

\begin{algorithm}[t]
\caption{Inertia-Informed FL Control}
\label{alg:FLC}
\scriptsize
\begin{algorithmic}[1]
\Require Inertia constants $\{H_i\}_{i \in \mathcal{G}}$; 
         weights $w_i = H_i / H_{\mathrm{total}}$
\State Initialize $\theta_i$ for all $i \in \mathcal{G}$
\While{\textbf{True}}
    \State \textit{//  Offline Centralized Training Phase }
    \For{each agent $G_{i},\ \forall i \in \mathcal{G}$, \textbf{in parallel}}
        \State Receive PMU data: $\omega_i$, $\Delta\delta_i$ at time $t$
        \State Compute RoCoF: $\dot{\omega}_i \gets [\omega_i(t) - \omega_i(t{-}\Delta t)] / \Delta t$
        \State Query centralized controller to obtain $P_{a,i}$
        \State Update $\theta_i$ via gradient descent on $\mathcal{L}_i(\theta_i)$ from \eqref{eqn:local_loss}
        \State Upload $\theta_i$ to FL server
    \EndFor
    \State \textit{//  Global Model Fusion }
    \State \textbf{Server:} $\theta_{G} \gets \sum_{i \in \mathcal{G}} w_i\,\theta_i$  from \eqref{eqn: IIWFedAvg}
    \State Broadcast $\theta_{G}$ to all agents
    \State \textit{//  Online Decentralized Execution Phase }
    \For{each agent $G_{i},\ \forall i \in \mathcal{G}$}
        \If{centralized control unavailable at bus $i$}
            \State $\hat{P}_{a,i} \gets f_{\theta_G}(\omega_i, \Delta\delta_i, \dot{\omega}_i, t)$
            \State $\hat{P}_{u,i} \gets -(\hat{P}_{a,i} - P_{d,i})$  from \eqref{eqn: FLC Pu}
        \EndIf
    \EndFor
\EndWhile
\end{algorithmic}
\end{algorithm}


\subsection{Performance Metrics}
\label{sec:metrics}

To quantify controller effectiveness and diagnose failure mechanisms under closed-loop deployment, we evaluate system performance using a combination of established stability indices and newly introduced metrics:

\subsubsection{\textbf{Frequency Nadir}}
the frequency nadir is a critical metric used to quantify the maximum frequency deviation following a disturbance \cite{hatziargyriou2021stability}. For each generator $i$, the nadir is defined as the maximum absolute frequency deviation from the nominal value over the simulation horizon $T_{\max}$:
\begin{equation}\label{eq:nadir}
\mathrm{Nadir}_i = \max_{t \in [0, T_{\max}]} |\Delta\omega_i(t)|.
\end{equation}
\subsubsection{\textbf{Stability Time}}
the stability time is defined as the elapsed time from the controller activation until the generator rotor speed deviation returns to, and remains within, the stability band $|\Delta\omega_i| \leq 0.01$~pu. This metric quantifies the duration of the transient period and the efficiency of the control action in restoring the quasi-steady state.
\subsubsection{\textbf{Destabilizing Fraction}}
we define the \emph{destabilizing fraction} $DF_i$ to quantify how often a controller at generator~$i$ injects power in a direction that amplifies, rather than damps, local frequency deviations. It is given by
\begin{equation}\label{eq:destab}
DF_i = \frac{1}{N_T}\sum_{k=1}^{N_T} \mathbf{1}\!\left[\mathrm{sign}(P_{u,i}(k)) = \mathrm{sign}(\dot{\omega}_i^{\mathrm{filt}}(k))\right] \times 100\%,
\end{equation}
where $\mathrm{sign}(\cdot)$ denotes the signum function, $N_T$ denotes the total number of discrete-time samples over the simulation horizon, and $\dot{\omega}_i^{\mathrm{filt}}$ is the filtered RoCoF signal, obtained via the first-order low-pass filter:
\[
\dot{\omega}_i^{\mathrm{filt}}(k) = (1-\alpha)\dot{\omega}_i^{\mathrm{filt}}(k-1) + \alpha\,\dot{\omega}_i(k),
\]
with $\alpha = \frac{\Delta t}{\tau + \Delta t}$, $\Delta t$ the sampling interval, and $\tau$ is the time constant. This filtering is essential to suppress high-frequency numerical noise: without it, the sign of $\dot{\omega}_i$ may fluctuate rapidly due to finite-difference and rounding errors, leading to spurious sign changes and unreliable estimates of destabilizing behavior. A stabilizing controller should oppose frequency deviations; thus, same-sign actions indicate destabilizing behavior. Values near $50\%$ correspond to ineffective (random) control, while values exceeding $50\%$ indicate systematic destabilization.

\subsubsection{\textbf{Topological Distance}}
to assess controller performance in relation to the power network topology, let the distance between a faulted bus $b_f$ and a generator bus $b_{G_i}$ be defined as the minimum number of transmission lines on any path in the bus graph \cite{pagani2013power}:
\begin{equation}\label{eq:elec_dist}
d(b_f,\, b_{G_i}) = \min_{\pi \in \Pi(b_f,\, b_{G_i})} |\pi|,
\end{equation}
computed via breadth-first search. This metric serves as a proxy for fault-induced perturbation severity at each generator: a smaller $d$ implies stronger electrical coupling to the fault and, consequently, a larger transient in the local state trajectory.

\begin{figure}[!t]
    \centering
    \includegraphics[page=1, width=0.68\columnwidth, trim=0mm 1mm 0mm 0.5mm, clip]{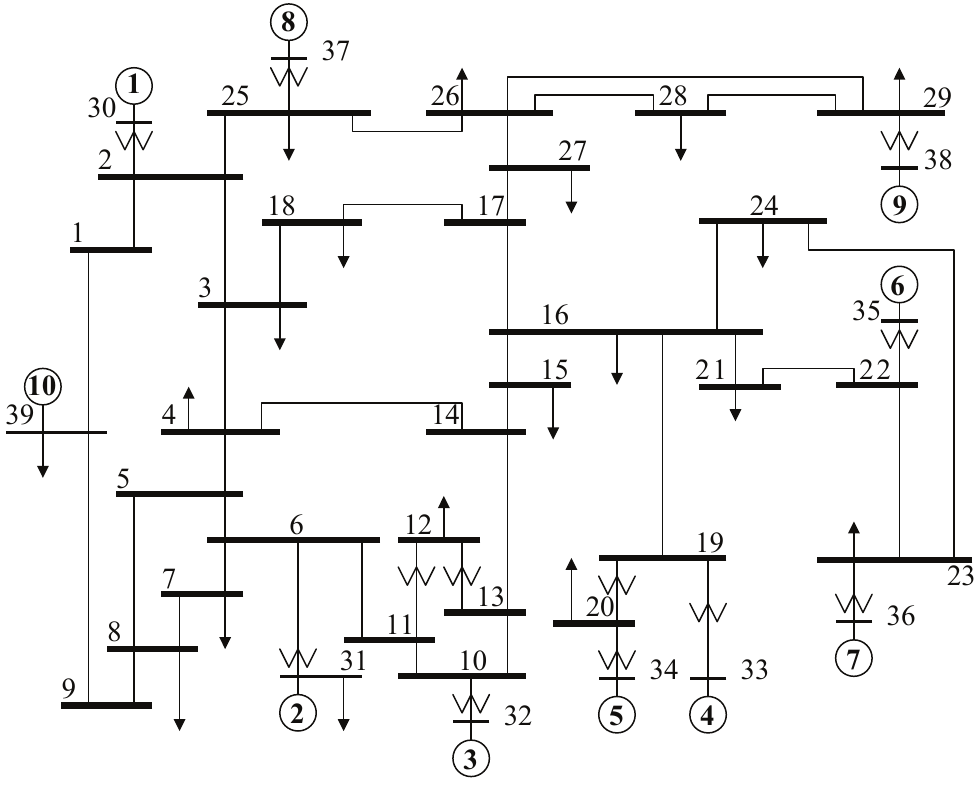}
    \caption{IEEE-39 New England bus system single-line diagram~\cite{IEEE_39}.}
    \label{fig:ieee39_sld}
\end{figure}

\section{Experimental Setup}

The FL framework is implemented in a client--server architecture using the Flower platform~\cite{flowerFL}. The ChebyKAN models adopt the architecture in~\cite{FLC_Paper} and are implemented via the Deep-KAN package~\cite{deepkan2024}. Training is conducted on MATLAB-generated transient stability data from the IEEE 39-bus, 10-generator benchmark system (Fig.~\ref{fig:ieee39_sld}), with subsequent closed-loop deployment in the same MATLAB environment. The proposed FL scheme achieves convergence within only 20 communication rounds, highlighting its high training efficiency and low communication overhead. All simulations and model training are performed on a Debian GNU/Linux virtual machine (kernel 6.1.0-33-amd64) with 8 CPU cores at 2~GHz and 64~GB RAM, using MATLAB R2024b for power system simulations. 
\begin{wraptable}[13]{r}{0.28\textwidth}
\vspace{-2pt}
\centering
\small{
\setlength{\tabcolsep}{2pt}
\renewcommand{\arraystretch}{0.8}
\caption{Generator inertia constants and IIWFedAvg weights.}
\label{tab:inertia}
\begin{tabular}{@{}c c c@{}}
\toprule
\textbf{Generator} & \textbf{$H_i$ (s)} & \textbf{$w_i$} \\
\midrule
$G_{10}$ & 42.0 & 0.1486 \\
$G_3$    & 35.8 & 0.1266 \\
$G_6$    & 34.8 & 0.1231 \\
$G_9$    & 34.5 & 0.1220 \\
$G_2$    & 30.3 & 0.1072 \\
$G_4$    & 28.6 & 0.1012 \\
$G_7$    & 26.4 & 0.0934 \\
$G_5$    & 26.0 & 0.0920 \\
$G_8$    & 24.3 & 0.0860 \\
\midrule
\multicolumn{2}{c}{$H_{\mathrm{total}}$} & 282.7\,s \\
\bottomrule
\end{tabular}
\vspace{6pt}
}
\end{wraptable}

Table~\ref{tab:inertia} reports the extracted inertia values and the corresponding IIWFedAvg weights for all active generators in the IEEE-39 Bus system~\cite{IEEE_39}.
The training data for each active generator $G_{i}$ bus represents input features $(\omega_i, \Delta\delta_i, \dot{\omega}_i$) at each time step $t$, and output label ($P_{a,i}$) corresponding to the computed centralized control action in \eqref{eqn: CPFL Pu} after a three-phase fault (F1) described in Table~\ref{tab:fault_details}. 
The fault is initiated at $t_{\text{fault}} = 0.5$ seconds and cleared at $t_{\text{clearing}} = 0.75$ seconds, with system trajectories sampled at $\Delta t = 10^{-2}$ seconds over a simulation horizon of $t_{\text{max}} = 1000$ seconds. The time constant for calculating $DF_{i}$ in the low-pass filter is set to $\tau=0.5$. In all experiments, the frequency and phase stabilization gains are fixed to $\alpha_i = 0.5$ and \mbox{$\beta_i = 0.005$}.

\begin{wraptable}[8]{r}{0.25\textwidth} 
\centering
\small{
\setlength{\tabcolsep}{2pt}    
\renewcommand{\arraystretch}{0.8} 
\caption{Fault Details}
\label{tab:fault_details}
\begin{tabular}{@{}c c c@{}}  
\toprule
\textbf{Fault} & \textbf{Faulted Bus} & \textbf{Tripped Line} \\
\midrule
F1 & 17 & 17--18 \\
F2 & 11 & 10--11 \\
F3 & 22 & 21--22 \\
F4 & 29 & 28--29 \\
F5 & 5  & 5--8   \\
\bottomrule
\end{tabular}
\vspace{6pt}
}
\end{wraptable}

Governor control is disabled throughout to test the designed control scheme's standalone distributed stabilization capability under unseen three-phase short-circuit faults (F2-F5), also detailed in Table~\ref{tab:fault_details}.

\section{Numerical Results \& Discussion}
\begin{figure*}[!t]
    \centering
    \begin{subfigure}{0.235\textwidth}
        \centering
        \includegraphics[width=\linewidth]{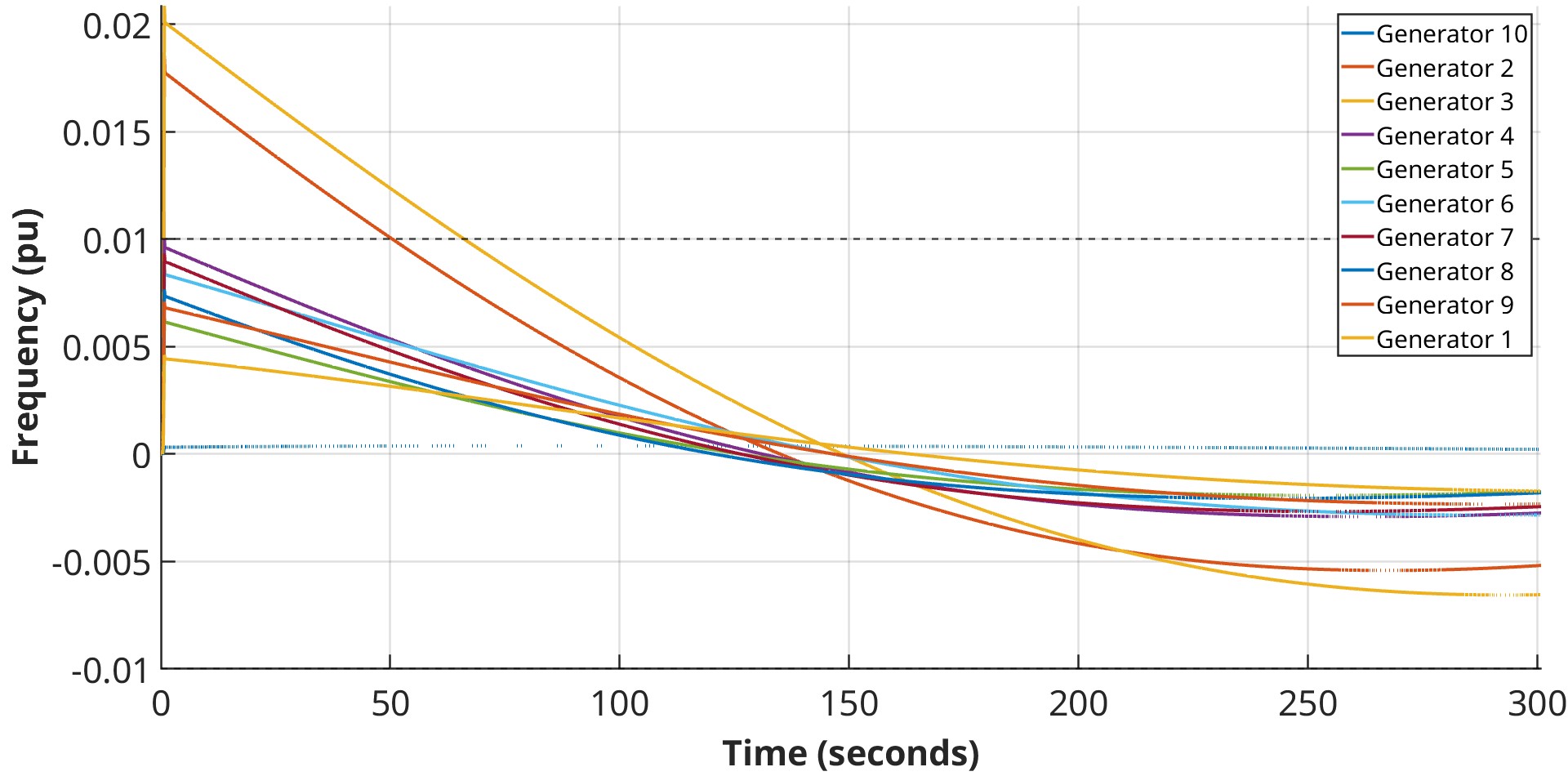}
        \caption{CPFL: angular frequency.}
        \label{fig:f2_freq_cpfl}
    \end{subfigure}\hfill
    \begin{subfigure}{0.235\textwidth}
        \centering
        \includegraphics[width=\linewidth]{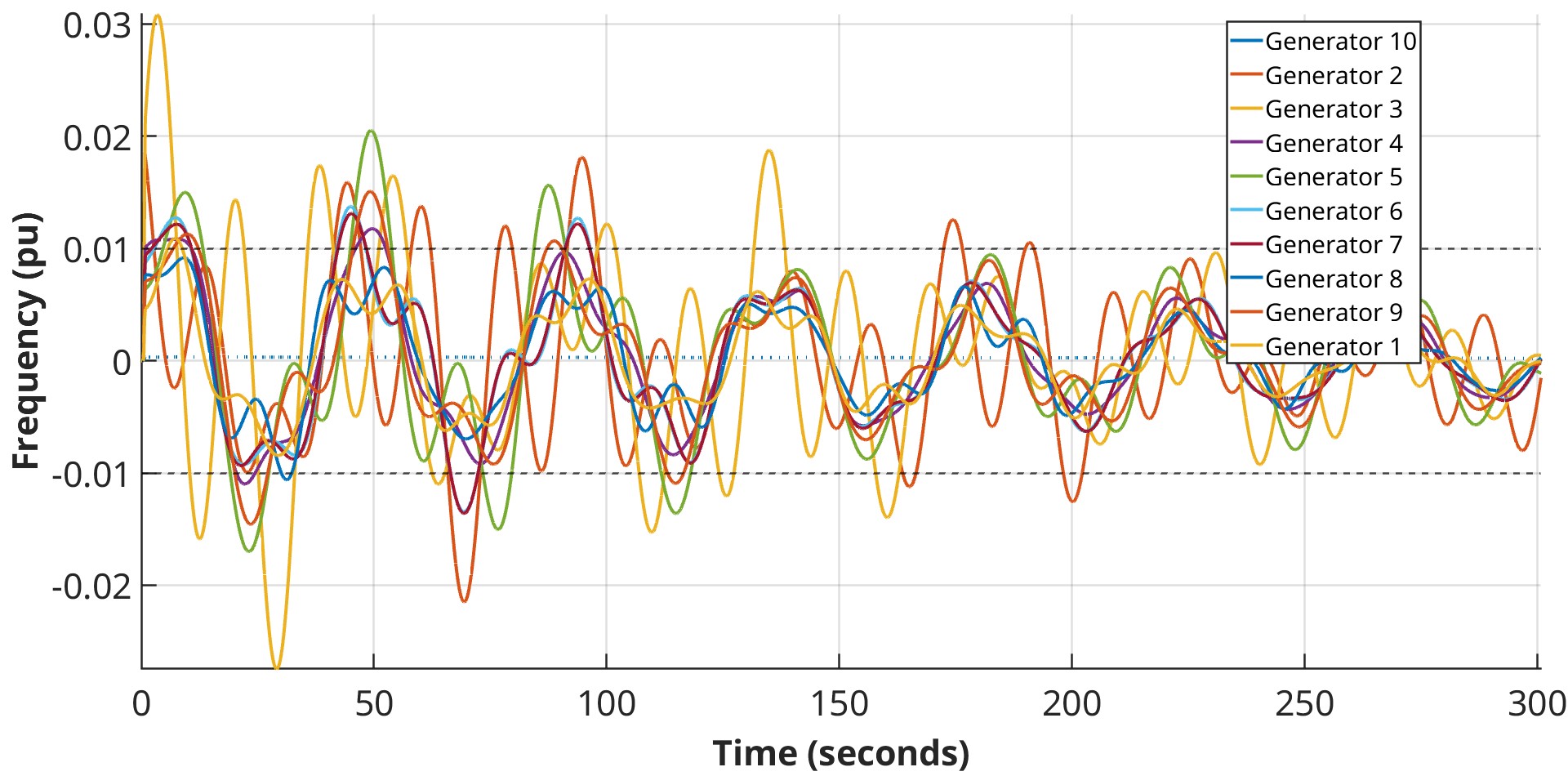}
        \caption{DPFL: angular frequency.}
        \label{fig:f2_freq_dpfl}
    \end{subfigure}\hfill
    \begin{subfigure}{0.235\textwidth}
        \centering
        \includegraphics[width=\linewidth]{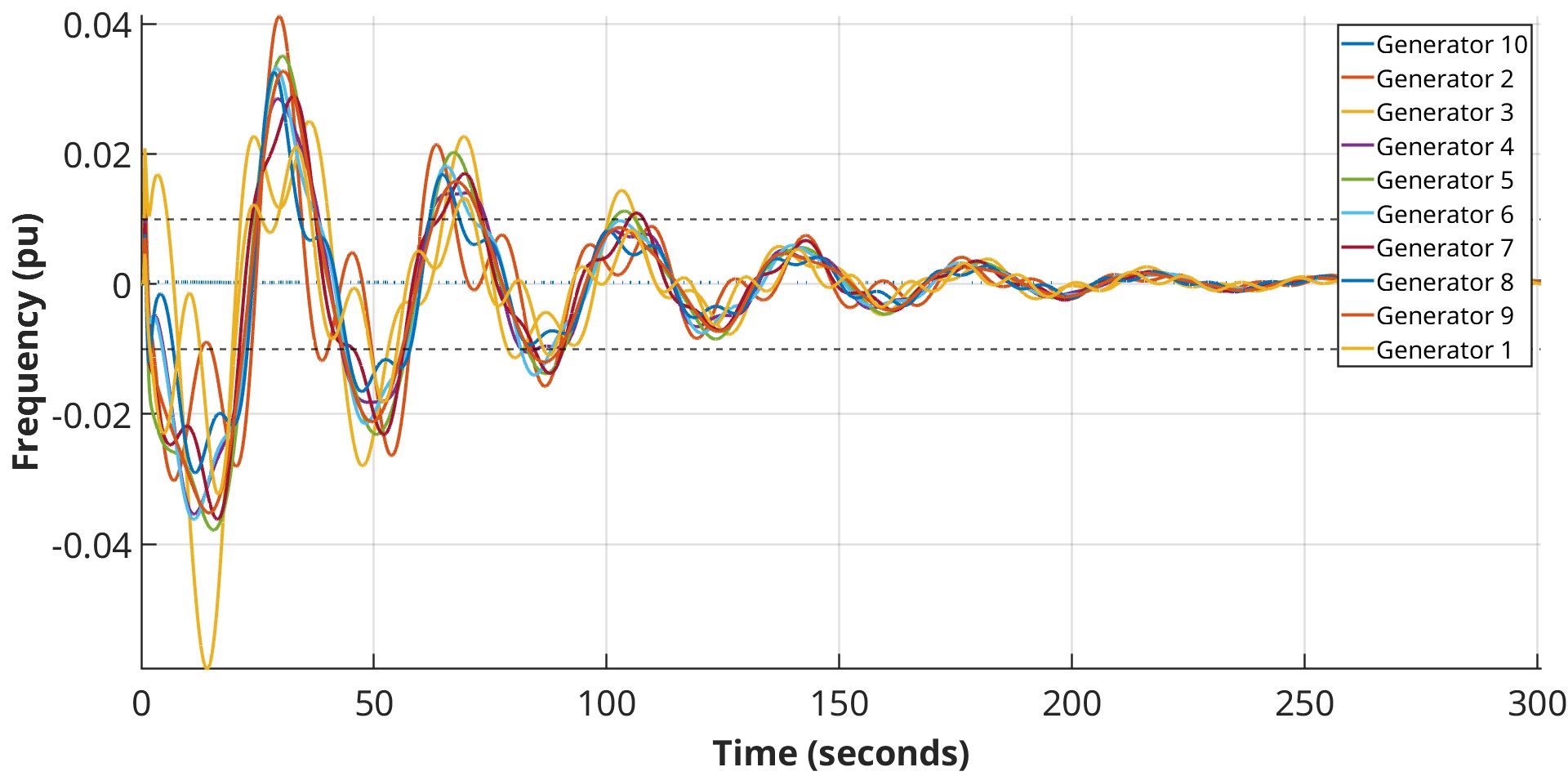}
        \caption{IIWFedAvg (w/o RoCoF): angular frequency.}
        \label{fig:f2_freq_no_rocof}
    \end{subfigure}\hfill
    \begin{subfigure}{0.235\textwidth}
        \centering
        \includegraphics[width=\linewidth]{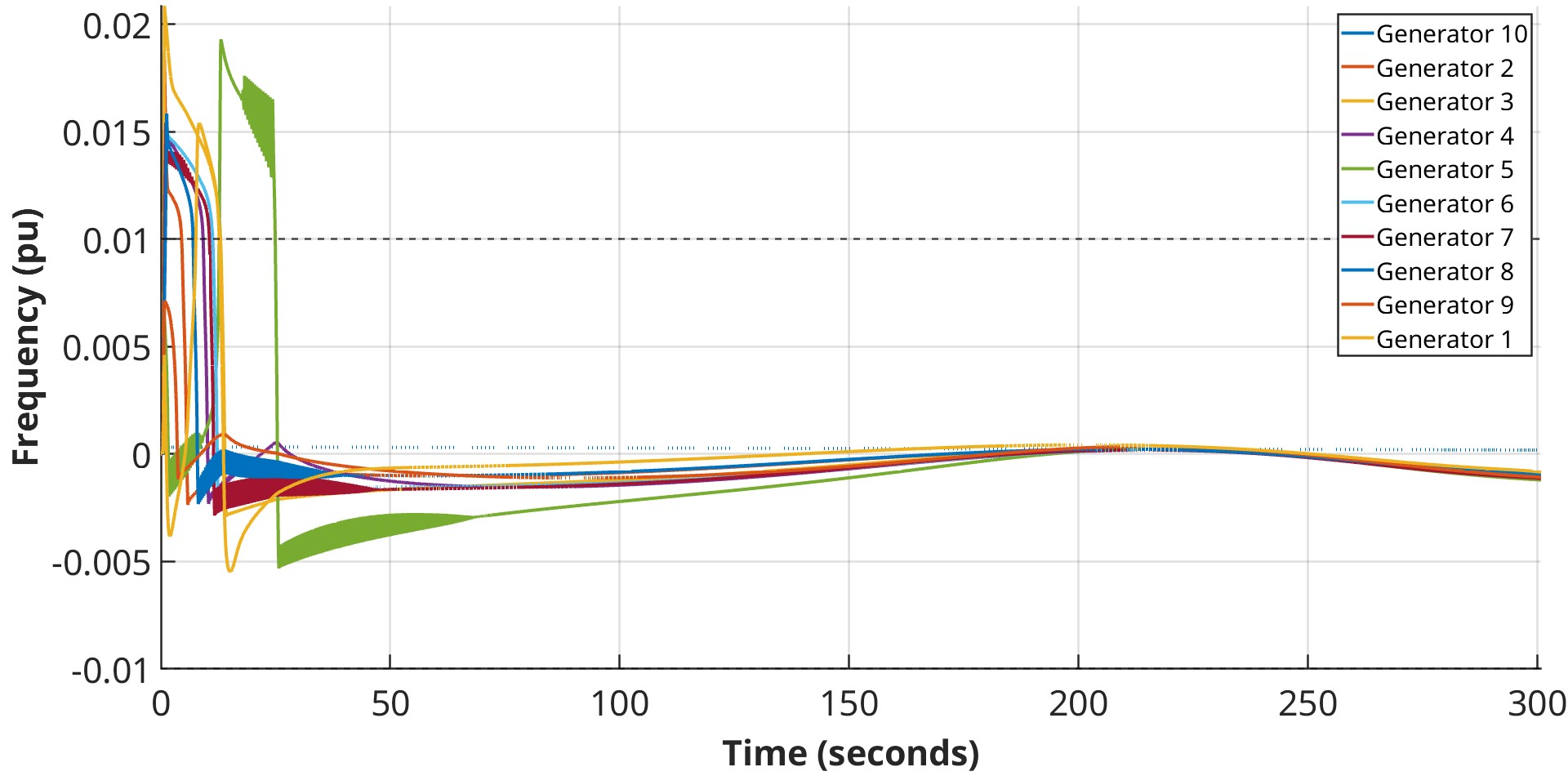}
        \caption{IIWFedAvg (w/ RoCoF): angular frequency.}
        \label{fig:f2_freq_rocof}
    \end{subfigure}

    \vspace{0.3em}

    \begin{subfigure}{0.235\textwidth}
        \centering
        \includegraphics[width=\linewidth]{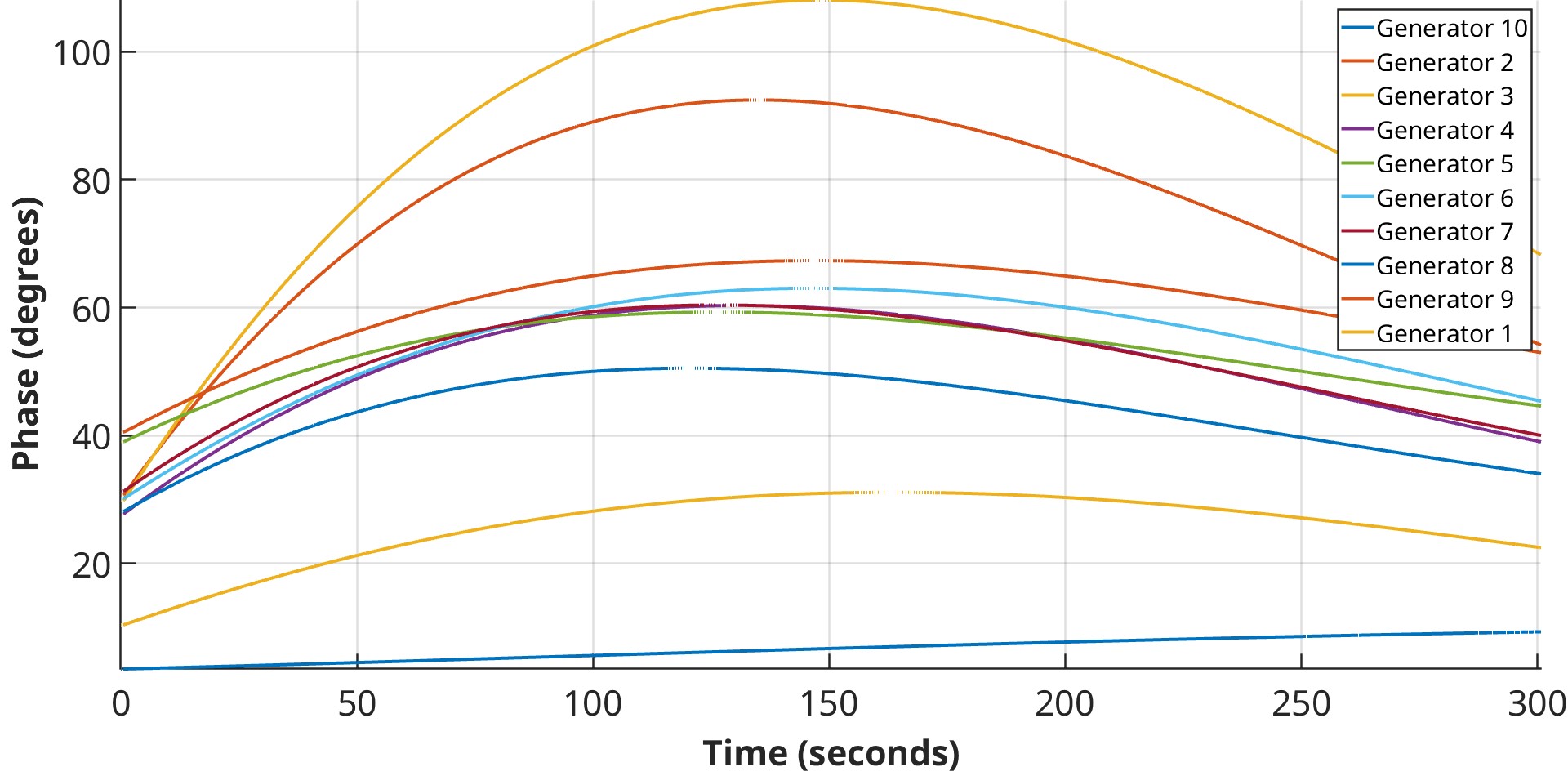}
        \caption{CPFL: rotor angle.}
        \label{fig:f2_phase_cpfl}
    \end{subfigure}\hfill
    \begin{subfigure}{0.235\textwidth}
        \centering
        \includegraphics[width=\linewidth]{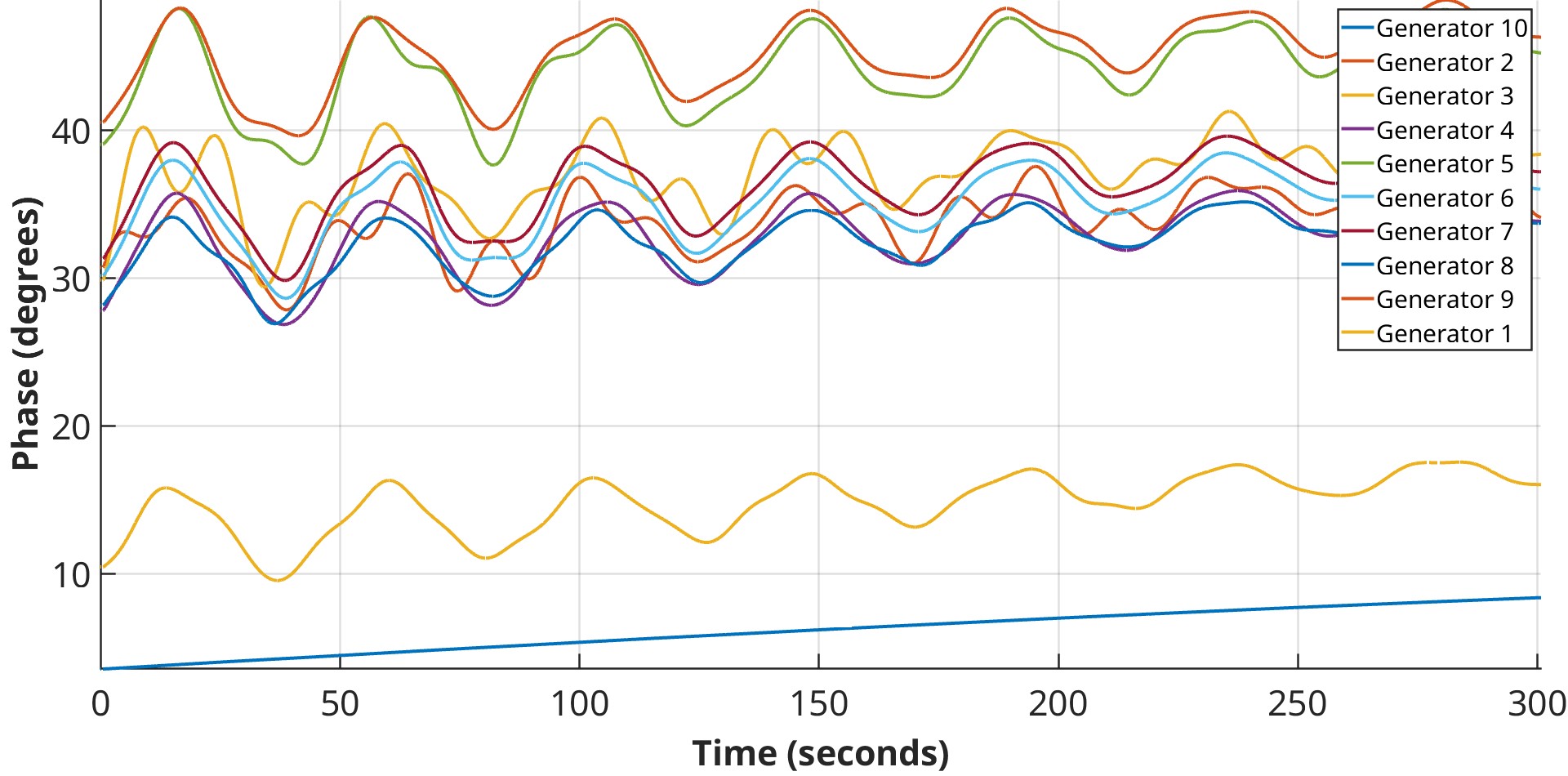}
        \caption{DPFL: rotor angle.}
        \label{fig:f2_phase_dpfl}
    \end{subfigure}\hfill
    \begin{subfigure}{0.235\textwidth}
        \centering
        \includegraphics[width=\linewidth]{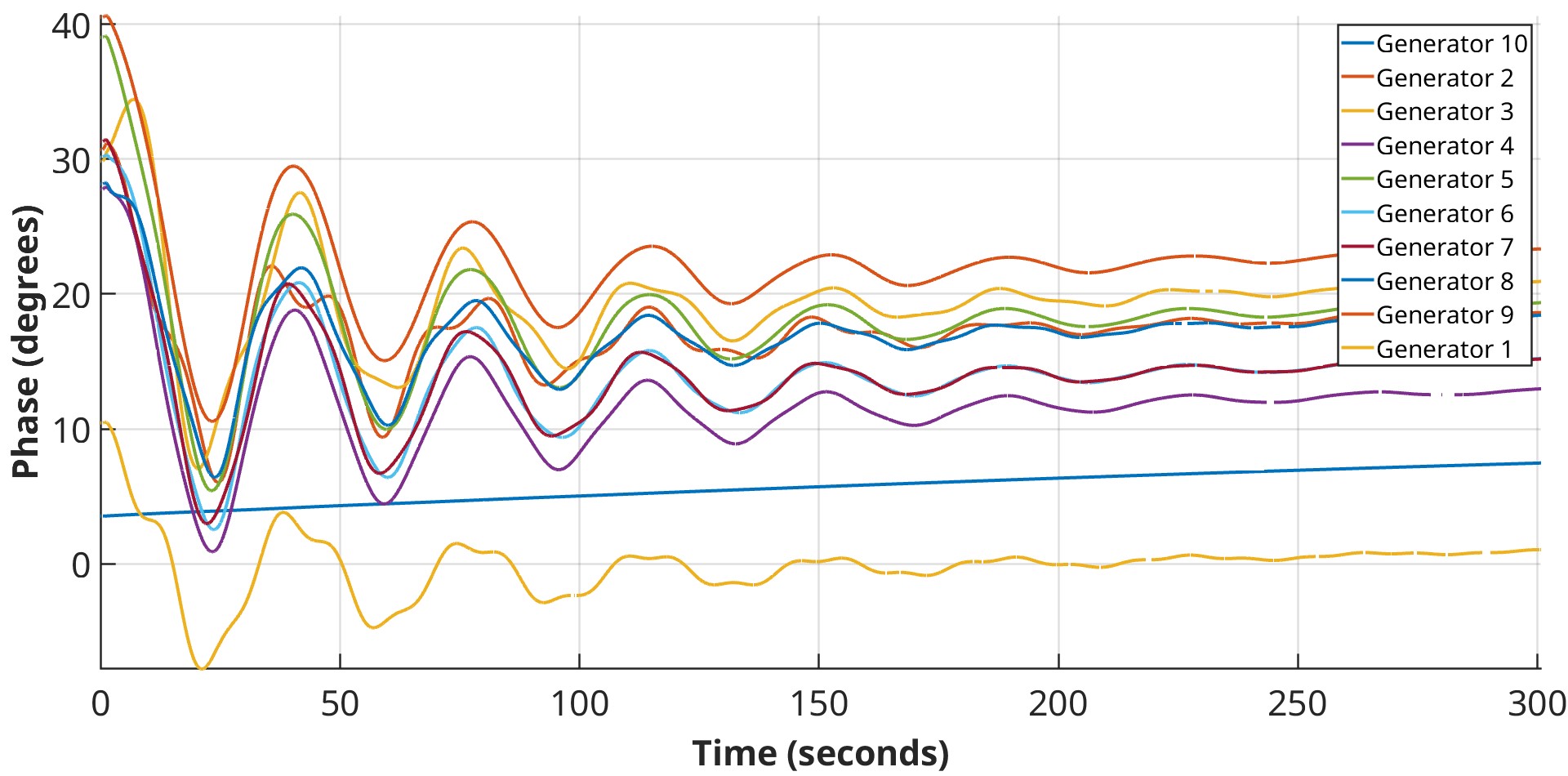}
        \caption{IIWFedAvg (w/o RoCoF): rotor angle.}
        \label{fig:f2_phase_no_rocof}
    \end{subfigure}\hfill
    \begin{subfigure}{0.235\textwidth}
        \centering
        \includegraphics[width=\linewidth]{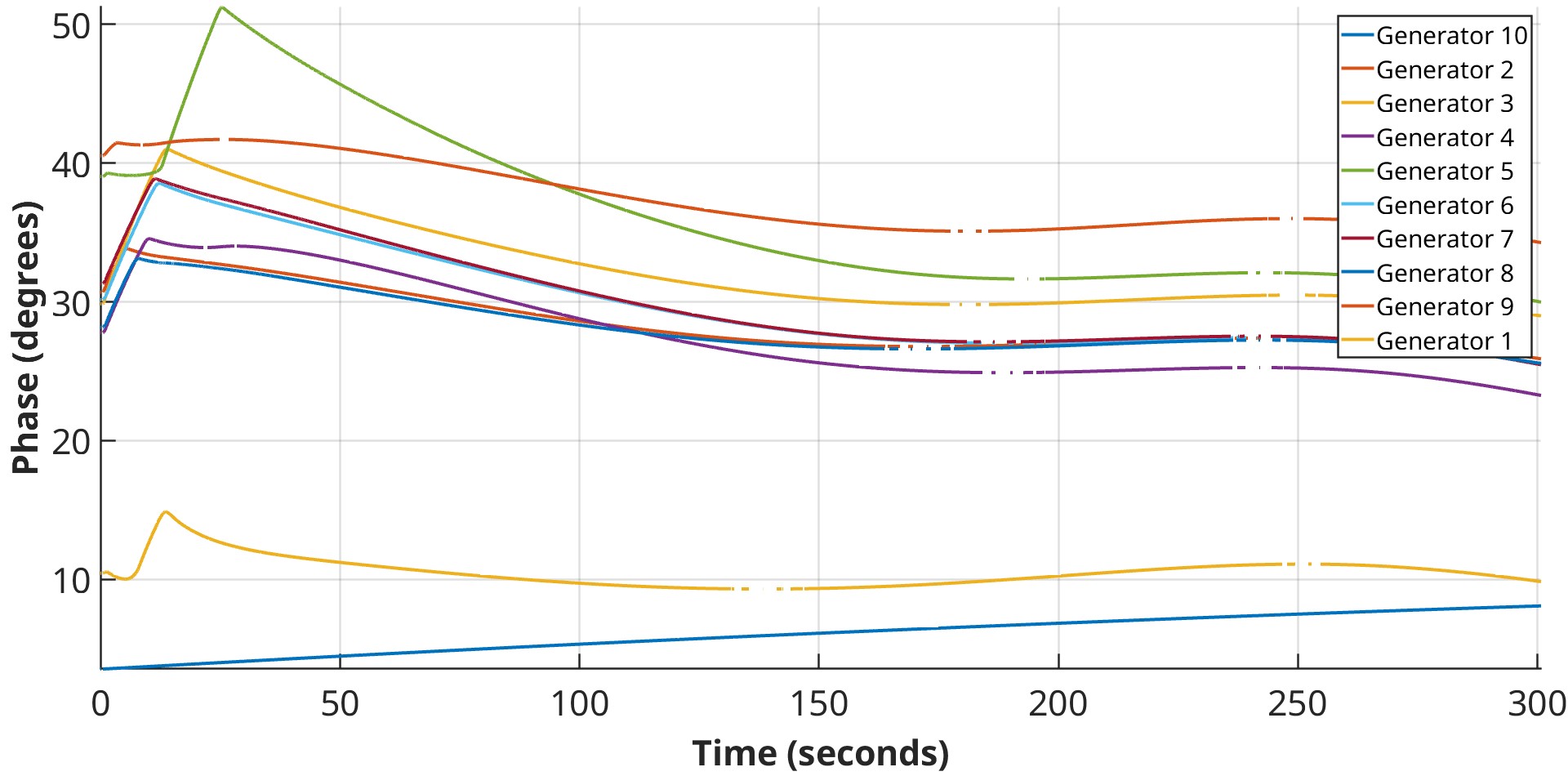}
        \caption{IIWFedAvg (w/ RoCoF): rotor angle.}
        \label{fig:f2_phase_rocof}
    \end{subfigure}

    \caption{Closed-loop frequency deviation and rotor angle trajectories under fault~F2.}
    \label{fig:f2_all}
\end{figure*}
\subsection{Generalization Under Closed-Loop Deployment}
To assess the generalization capability of the proposed IIWFedAvg controller, two ChebyKAN model variants are trained exclusively on fault~F1 and are deployed and tested for closed-loop stabilization performance under four unseen three-phase fault contingencies~(F2--F5) at full decentralized deployment, i.e., generator busses ~$G_2$--$G_{10}$ are governed by the IIWFedAvg neural controllers while~$G_1$ retains conventional control. Performance is benchmarked against the CPFL controller with full-system PMU observations and the DPFL baseline using local PMU observations. Table~\ref{tab:piwavg_results} reports the average stability times across all active generators, where~$\infty$ indicates that system-wide stability was not achieved within a 0-300-second simulation execution window.
\begin{table}[!h]
\centering
\scriptsize
\caption{Average stability times (seconds) under unseen fault scenarios for CPFL, DPFL, and the two IIWFedAvg variants at 100\% decentralized deployment.}
\label{tab:piwavg_results}
\setlength{\tabcolsep}{4pt}
\renewcommand{\arraystretch}{1.15}
\begin{tabular}{c|c|c|cc}
\toprule
\multirow{2}{*}{\textbf{Fault}} & \multirow{2}{*}{\textbf{CPFL}} & \multirow{2}{*}{\textbf{DPFL}} & \multicolumn{2}{c}{\textbf{IIWFedAvg}} \\
& & & \textbf{w/o RoCoF} & \textbf{w/ RoCoF} \\
\midrule
F2 & 11.5 & 87.9 & 82.4 & \textbf{8.7} \\
F3 & 14.9 & $\infty$ & $\infty$ & $\infty$ \\
F4 & 8.7  & $\infty$ & 64.7 & \textbf{2.9} \\
F5 & 12.5 & 75.8 & 85.4 & \textbf{38.9} \\
\midrule
Stabilized & 4/4 & 2/4 & \textbf{3/4} & \textbf{3/4} \\
\bottomrule
\end{tabular}
\end{table}

The IIWFedAvg controller achieves system-wide stabilization under fully decentralized operation for three of the four unseen faults, yielding a generalization success rate of 75\% despite training on a single fault distribution. This demonstrates that the inertia-informed federated aggregation strategy enables the learned controller to transfer effectively across unseen fault scenarios at full decentralized deployment. 

RoCoF augmentation consistently accelerates transient recovery. Under F2, it achieves a $10\times$ speedup with respect to average system stability time over DPFL and enables stabilization under F4, where DPFL fails entirely. While IIWFedAvg without RoCoF achieves stability, it exhibits slower damping, confirming that RoCoF is critical for rapid stabilization rather than stability alone. Most notably, the RoCoF-augmented IIWFedAvg controller surpasses the centralized CPFL benchmark on two of three stabilized faults (F2 \& F4), achieving a~$3\times$ speedup under~F4 at full decentralized deployment. This outcome inverts the conventional centralized--decentralized performance trade-off: IIWFedAvg enabled fully distributed deployment by incorporating the inertia factor in federated aggregation strategy, while RoCoF augmentation closes the observability gap by serving as a communication-free proxy for the unobserved global states.

Fig.~\ref{fig:f2_all} reveals a clear performance hierarchy under fault~F2. Inertia-informed aggregation alone does not accelerate damping. RoCoF augmentation is the decisive factor, driving generator angular frequency deviations to within~$\pm0.01$\,pu, faster than the centralized CPFL benchmark without inter-agent communication. Rotor angle synchronism is preserved across all controllers~(Figs.~\ref{fig:f2_phase_cpfl}--\ref{fig:f2_phase_rocof}), confirming that the performance gain under F2 is attributed to damping speed rather than synchronism recovery.

\subsection{Limitations Assessment}

In the considered case study, the proposed control architecture showed limited gains against one fault (F3) in particular. To determine whether this originated from insufficient training data or from the proposed aggregation strategy, IIWFedAvg models are trained independently for each fault (F1--F5) and tested on F3. Table~\ref{tab:f3_gen} reports the topological distance from each generator bus to each faulted bus, eq.~\eqref{eq:elec_dist}, sorted by proximity to~F3. The bottom rows aggregate the IIWFedAvg weight budget (Table~\ref{tab:inertia}) by proximity zone: $W_{\textsc{close}}$ sums the weights of generators within 2~hops of the fault, $W_{\textsc{mod}}$ covers 3--4~hops, and $W_{\textsc{far}}$ covers $\geq$5~hops. For~F3, 68.2\% of the aggregation weight resides in the far zone, while the two generators adjacent to the fault ($G_6$, $G_7$) hold only 21.6\%.
\begin{table}[!t]
\centering
\scriptsize
\caption{Topological distance (hops) from each generator to each fault bus, sorted by proximity to~F3.}
\label{tab:f3_gen}
\setlength{\tabcolsep}{4pt}
\renewcommand{\arraystretch}{1.1}
\begin{tabular}{c|ccccc}
\toprule
\textbf{Gen.} & \textbf{F1} & \textbf{F2} & \textbf{F3} & \textbf{F4} & \textbf{F5} \\
\midrule
$G_6$    & 4 & 8 & \textbf{1} & 7 & 7 \\
$G_7$    & 4 & 8 & 2 & 7 & 7 \\
$G_4$    & \textbf{3} & 7 & 4 & 6 & 6 \\
$G_5$    & 4 & 8 & 5 & 7 & 7 \\
$G_3$    & 6 & \textbf{2} & 7 & 9 & 4 \\
$G_8$    & 4 & 7 & 7 & 3 & 5 \\
$G_9$    & 4 & 9 & 7 & \textbf{1} & 7 \\
$G_{10}$ & 4 & 6 & 7 & 4 & 4 \\
$G_2$    & 6 & \textbf{2} & 8 & 8 & \textbf{2} \\
\midrule
$W_{\textsc{close}}$ ($\leq$2) & 0.0\% & 23.4\% & 21.6\% & 12.2\% & 10.7\% \\
$W_{\textsc{mod}}$ (3--4) & 76.6\% & 0.0\% & 10.1\% & 23.5\% & 27.5\% \\
$W_{\textsc{far}}$ ($\geq$5) & 23.4\% & 76.6\% & 68.2\% & 64.3\% & 61.8\% \\
\bottomrule
\end{tabular}
\end{table}

\begin{table}[!t]
\centering
\scriptsize
\caption{F3 diagnostic: top-2 generators by nadir and by destabilizing fraction~$DF_i$ for each training split.}
\label{tab:f3_diag}
\setlength{\tabcolsep}{2pt}
\renewcommand{\arraystretch}{1.1}
\begin{tabular}{c|ccc|ccc|ccc|ccc}
\toprule
 & \multicolumn{6}{c|}{\textbf{Frequency Nadir}} & \multicolumn{6}{c}{\textbf{Destabilizing Fraction }} \\
 & \multicolumn{3}{c|}{\textbf{1st}} & \multicolumn{3}{c|}{\textbf{2nd}} & \multicolumn{3}{c|}{\textbf{1st}} & \multicolumn{3}{c}{\textbf{2nd}} \\
\textbf{Train} & \textbf{Gen} & \textbf{pu} & \textbf{hops} & \textbf{Gen} & \textbf{pu} & \textbf{hops} & \textbf{Gen} & \textbf{$DF_i$} & \textbf{hops} & \textbf{Gen} & \textbf{$DF_i$} & \textbf{hops} \\
\midrule
F1 & $G_2$ & 3.5 & 8 & $G_3$ & 3.4 & 7 & $G_7$ & 64.7 & \textbf{2} & $G_6$ & 60.1 & \textbf{1} \\
F2 & $G_9$ & 2.17 & 7 & $G_8$ & 2.13 & 7 & $G_7$ & 79.9 & \textbf{2} & $G_6$ & 73.4 & \textbf{1} \\
\textbf{F3} & $\bm{G_{10}}$ & \textbf{2.69} & \textbf{7} & $\bm{G_5}$ & \textbf{2.66} & \textbf{5} & $\bm{G_5}$ & \textbf{65.7} & \textbf{5} & $\bm{G_9}$ & \textbf{63.3} & \textbf{7} \\
F4 & $G_3$ & 3.9 & 7 & $G_2$ & 3.8 & 8 & $G_7$ & 72.4 & \textbf{2} & $G_6$ & 69.6 & \textbf{1} \\
F5 & $G_9$ & 2.8 & 7 & $G_8$ & 2.8 & 5 & $G_7$ & 71.7 & \textbf{2} & $G_6$ & 70.8 & \textbf{1} \\
\bottomrule
\end{tabular}
\end{table}

Table~\ref{tab:f3_diag} reports the two generators with the highest nadir and the two with the highest destabilizing fraction~$DF_i$ for each training split, along with their hop distance to~F3. All five controllers fail to stabilize the system, including the in-distribution case (trained and tested on~F3). Two consistent patterns emerge across all training splits. First, the highest nadirs always occur at far-field generators (5--8~hops), indicating that remote generators may experience the largest frequency excursions. Second, across all four out-of-distribution splits, the highest destabilizing fractions are observed for $G_7$ and $G_6$ (1--2~hops), with $DF_i$ values of 60--80\%, indicating that the controllers closest to the fault inject power in the wrong direction most of the time. When the controller is trained on~F3 itself, $G_6$ and $G_7$ are no longer the top destabilizers, confirming that their local models can learn the correct dynamics, but the aggregation dilutes them with 68.2\% far-field weight concentration, shifting the destabilizing burden to remote generators.

\section{Conclusion}

Collaborative FL-based CTDE control framework in smart grids demonstrated generalization capabilities to unseen disturbances, yet standard FL aggregation strategies remain agnostic to the physical heterogeneity of synchronous generators. This work proposed Inertia-Informed Weighted FedAvg (IIWFedAvg), a physics-informed FL aggregation strategy that embeds generator inertia directly into global model fusion, enabling fully decentralized transient stability control in smart grid transmission networks. Furthermore, RoCoF feature augmentation enhances interpretable ChebyKAN neural controllers' actions by incorporating an implicit proxy for unobserved global dynamics, computed from local PMU measurements without centralized coordination overhead.

IIWFedAvg is evaluated on the IEEE 39-bus benchmark under full decentralized deployment. IIWFedAvg achieves a 75\% generalization success rate across four unseen fault contingencies, despite training on a single fault distribution. Notably, the RoCoF-augmented variant surpasses the CPFL benchmark in two out of three stabilized faults and achieves a 3× speedup under F4 with zero inter-agent communication cost. Together, these 
results demonstrate that physics-informed aggregation and local feature augmentation can collectively close the centralized--decentralized performance gap without any coordination overhead.

Diagnostic analysis of the F3 failure indicates that topology-blind aggregation induces two coupled effects: far-field generators experience the largest frequency deviations, while near-fault generators exhibit the highest destabilizing fractions. This pattern persists across training distributions, suggesting that inertia-based weighting is not sufficiently correlated with fault proximity. Consequently, near-fault policies critical for stabilization are systematically attenuated, highlighting the need for topology-aware aggregation as a key direction for future work.

\bibliographystyle{IEEEtran}
\vspace{-2mm}
\bibliography{references_clean}

\end{document}